\documentstyle{mn}
\input{epsf}
\newcommand{\be}{\begin{eqnarray}}
\newcommand{\ee}{\end{eqnarray}}
\newcommand{\hMpc}{{\ifmmode{h^{-1}{\rm Mpc}}
\else{$h^{-1}$Mpc}\fi}}

\title[Effect of small-scale perturbations]
{Effect of small scale density perturbations on the formation of dark matter halo profiles}
\author[Pilipenko, Doroshkevich, Lukash, Mikheeva]
{S.V. Pilipenko$^{1,2}$, A.G. Doroshkevich$^1$, V.N. Lukash$^1$, E.V. Mikheeva$^1$\\
$1$Astro Space centre of Lebedev Physical
           Institute of  Russian Academy of Sciences, Profsojuznaja st. 84/32,
                        117997 Moscow,  Russia\\
        $2$ Moscow Institute of Physics and Technology, Institutskij per. 9, 
            141700 Dolgoprudnyj, Russia\\}
\date{Accepted ....,
      Received ...,
        in original form ... .}

\begin{document}
\maketitle
\begin{abstract}
With help of a set of toy N-body models of dark halo formation we study the impact of small scale initial perturbations on the inner density profiles of haloes. We find a significant flattening of the inner slope $\alpha={\mathrm{d} \log \rho \over \mathrm{d} \log r}$ to $\alpha=-0.5$ in some range of scales and amplitudes of the perturbations (while in the case of absence of these perturbations the NFW profile with $\alpha=-1$ is reproduced). This effect may be responsible for the formation of cuspless galactic haloes.
\end{abstract}

\section{Introduction}
The ``cusp problem'' is one of the most serious problems of the standard ($\Lambda$CDM) cosmology (e.g. Burkert 1995; de Blok 2001, 2010; Wyse \& Gilmore 2008). It is manifested as the existence of cuspless galaxies in the real Universe as found from observations of rotation curves of LSB galaxies. On the other hand cosmological N-body simulations show that all stable haloes have similar NFW-like cuspy profiles. A number of explanations have been proposed, including modifications of the $\Lambda$CDM model, such as warm dark matter (Tremaine \& Gunn 1979; Avila-Reese et al. 2001; Walker 2009), collisional DM (Spergel \& Steinhardt 2001), decaying DM (Cen 2001; Abdelqader 2008; Pilipenko et al. 2009), mixed (oscillating) DM particles (Medvedev 2012).

The cusp problem is a part of a more general physical problem: what is the connection of halo properties with the initial conditions (the field of cosmological density perturbations). Mikheeva, Doroshkevich \& Lukash (2007), Lukash \& Mikheeva (2010), Doroshkevich, Lukash \& Mikheeva (2012, hereafter DLM) proposed an analytical solution for this problem. Their idea is called an entropy theory and it is based on the calculation of the initial coarse grained entropy function\footnote{For an ideal gas the entropy $S\propto \log E$, so $E$ is called the entropy function.} profile of a protohalo, which is defined as
\be
E = \sigma^2 \rho^{-2/3},
\label{entrop}
\ee
where $\sigma$ is the velocity dispersion of DM particles in a protohalo and $\rho$ is their density as functions of a mass $M$ within a given radius $r$. This characteristic is connected with the well known coarse grained phase space density (Antonov 1961, Lynden-Bell 1967):
\[
 f = \rho \sigma^{-3} = E^{-3/2},
\]
which decreases with time during relaxation, hence the entropy function of a protohalo grows with time.
In DLM the entropy function is calculated at two moments of time: the first one is related to the moment of a protohalo collapse extrapolated from the linear theory while the second one corresponds to the epoch when the halo is already relaxed.

The central cusp has very low entropy, which tends to zero when $r\rightarrow0$ as fast as $E\propto M^{5/6}(\propto r^{5/3})$ if one assumes the $r^{-1}$ slope of the density profile. If the initial entropy of the central part is more shallow, the cusp will be suppressed at some radius which depends on an amplitude of the initial entropy. There are three possible sources of initial entropy: thermal motions of particles, which are very small in the case of CDM, small scale perturbations (on a scale much smaller than the mass of the inner part of a halo) and a complex shape of a protohalo. Since N-body simulations have limited resolution, the small scale part of the spectrum is always cut off, and therefore the initial entropy in simulations is always underestimated.

The effect of small scale density perturbations has been studied numerically in several papers, however the question of their effect on the process of halo formation remained unfold. For example, in Bagla et al. (2005) the impact of small scale fluctuations on the pancake formation was considered, Bagla \& Prasad (2009) focused on the effect of non-linear small scale perturbations on the large scale power spectrum. An interesting result was obtained in Ma \& Boylan-Kolchin (2004). They simulated equilibrium dark matter haloes with several subhaloes and found that subhaloes can flatten density profiles of host haloes.

In order to check the theory of DLM numerically and estimate the amplitude of the effect we study the impact of small scale perturbations on the formation of dark matter haloes using a simple toy model of three collapsing waves of deformation directed along three orthogonal axes. This model is known as one of simplest models where a NFW-like halo forms (Shapiro et al. 2004). It also has an attractive property that the collapse of one wave is an exact solution at the Zel'dovich theory (Zel'dovich 1970).

We add waves with wavelength 10-20 times smaller than that of the main three waves in order to reproduce the small scale perturbations. Our simulations show that this indeed results in a flattening of halo profile.

In this Letter we describe the initial conditions in Section 2, present the resulting halo profiles and compare them with the DLM theory in Section 3, discuss the connection with the realistic initial conditions and observations in Section 4 and summarize our conclusions in Section 5.

\section{Simulations and initial conditions}
The simulations are performed with the GADGET-2 code (Springel 2005). The periodic boundaries and comoving integration are turned on, the concordance $\Lambda$CDM parameters are used (the dimensionless density parameters $\Omega_{\rm m}=0.3$ and $\Omega_\Lambda=0.7$). However, the formation of the inner part of halo is chosen to take place at a high redshift $z>1$, so the impact of the $\Lambda$ term is negligible.

In the simplest case the initial conditions correspond to three sine waves of particle displacements from a Cartesian grid. The particle comoving positions $x^i_k$ and velocities $v^i_k$ are:
\begin{eqnarray}
\label{init}
x_k^i= \ell(i-1/2)+ A_k \sin \left({2\pi \ell(i-1/2) \over L} \right),\\
v_k^i= a_0 H(a_0) A_k \sin \left({2\pi \ell(i-1/2) \over L} \right), \nonumber
\end{eqnarray}
\[
\ell\equiv L/n;\, k=1,2,3;\,i=1,...,n,
\]
where $L$ is the comoving wavelength and simultaneously the box size, $n$ is the number of particles in a row,
\[
 A_k = {a_0 \over a_k}{L \over 2\pi}
\]
is an amplitude of $k$-th wave, $a_0$ is an initial time (we use the scale factor $a$ as a measure of time, $a=1$ at present), $a_k$ is the collapse time in the Zel'dovich approximation and $H(a_0)$ is the Hubble constant at $a_0$. The first term in the first equation in (\ref{init}) represents the Cartesian grid while the second one characterizes the displacement. 

We take the relative amplitudes 
\be
\label{normal}
A_2=A_3=0.2A_1.
\ee
The influence of relative amplitudes is discussed in Section 3. In this case the collapse times are taken $a_1=0.1$, $a_2=a_3=0.5$ respectively. They are small enough to neglect the effects of $\Lambda$-term. The initial time is chosen to be $a_0 = 1/80$.

\begin{figure}
\centering
\epsfbox{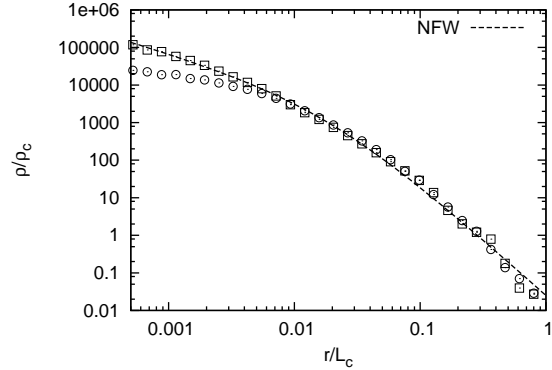}
\caption{The density profiles of dark matter haloes. The halo with no initial small scale perturbations is shown by squares, the halo with perturbations is shown by circles. The dashed line corresponds to the best-fitting NFW model with parameters given in eq. (\ref{fit}).}
\label{fig_profiles}
\end{figure}

The small scale modes have been added by introducing additional displacement term of the same form as in (\ref{init}) with a wavelength $\lambda$ and a collapse time $a_{\rm s}$. Their amplitudes along three directions are assumed to be equal to each other and written as:
\begin{eqnarray}
\label{small}
\Delta x_k^i=  {a_0 \over a_{\rm s}}{\lambda \over 2\pi} \sin \left({2\pi \ell(i-1/2) \over \lambda} \right),\\
\Delta v_k^i= a_0 H(a_0) {a_0 \over a_{\rm s}}{\lambda \over 2\pi} \sin \left({2\pi \ell(i-1/2) \over \lambda} \right). \nonumber
\end{eqnarray}
Although additional phases $\phi_k$ can be added to equations (\ref{small}), however our tests have shown that changing phases from 0 to $\pi/2$ does not change the profiles of haloes.

We have varied the resolution of our simulations from $8^3$ to $256^3$ particles. In all cases the final profile converges at scales larger than the gravitational softening length. Valuable results for the study of small scale effects can be achieved in $128^3$ and $256^3$ simulations. The latter has been chosen for a further study due to the best stability and precision. With such resolution the final halo has about $8\cdot 10^6$ particles inside virial radius. For lower resolution models with $64^3$ particles the halo consists of $10^5$ particles and the impact of small scale waves on the density profile is comparable to the numerical noise.

The comoving softening scale of the gravitational force (the $\epsilon$ value, which is roughly equivalent to the Plummer softening length, Springel 2005) for $128^3$-particle and $256^3$-particle simulations is chosen to be
\[
\epsilon = min(10.0, 0.75/a) \cdot 10^{-4} L.
\]
It is taken to be two times larger for the $64^3$-particle simulations. With such a choice the mean distance between the particles in the inner part of haloes is smaller than $\epsilon$ since the time of the first wave collapse.
Additional test runs of simulations with different softenings have shown that this amount of softening provides stable results.

We have searched for a highest density peak and have identified it with the halo centre for each simulated halo.

\section{Results and discussion}
First we reproduce the results of Shapiro et al. (2004) with a higher resolution ($256^3$ particles instead of $64^3$). A stable halo forms soon after the first collapse of the wave with the largest amplitude. However, the accretion continues at any time due to the choice of initial conditions and this changes the profile of the halo periphery. The density profile of a stable halo obtained from a $256^3$ particles simulation of three plane waves collapse is plotted in Figure \ref{fig_profiles} by squares. It is clear from this Figure that the NFW profile fits the data quite well. The statistical error due to the finite number of particles is much smaller than the symbol size on this plot, and the time variations of profiles in the central part ($r/L_{\rm c}<0.1$) are also smaller than or comparable to the plotting symbol size (but larger than the statistical error).

We fit the profile with the NFW formula (Navarro et al., 1995). The best-fitting parameters are:
\[
\rho(r) = {\rho_{\rm c} \over x \left( 1 + x \right)^2}, \quad 
x = {r \over r_{\rm c}}
\]
\begin{equation}
r_{\rm c} = (0.019\pm 0.001) L_{\rm c}, \quad \rho_{\rm c} = (3.8\pm 0.1)\cdot 10^3 \rho_{\rm m}.
\label{fit}
\end{equation}
Here as well as in Figure \ref{fig_profiles} $L_{\rm c} = L a_{\rm c}$ is the physical size of the simulation box (and the wavelength) at the time $a_{\rm c}$ of stabilization of the central part of the halo. $\rho_{\rm m}$ is the mean cosmological density of matter at that time. According to the definition of $a_{\rm c}$ the density profile at $a_{\rm c}$ differs less than 50\% from the final profile at each bin within $r<r_{\rm c}$ . For the adopted choice of parameters, (\ref{normal}), this time corresponds to the scale factor $a_{\rm c}=0.19=1.9a_1$.

We have varied the relative amplitudes of three waves and have found that this leads only to a change of the time of halo stabilization $a_{\rm c}$, while the shape of the profiles remains almost the same. If the profiles are plotted in the same coordinates as in Figure \ref{fig_profiles} (with its own $a_{\rm c}$ for each halo), the best-fitting scale parameter $r_{\rm c}$ varies only slightly and the $\rho_{\rm c}$ more prominent %parameters also vary only slightly 
with the variations of relative amplitudes. The dependence of halo central part stabilization time $a_{\rm c}$ on the relative amplitudes of three waves is shown in Table \ref{tab1}.

\begin{table}
\caption{The central part formation time and NFW profile fitting parameters for haloes with different initial relative amplitudes of three waves. Here $A_{1,2,3}$ are the amplitudes of three waves, $a_{\rm c}$ is the moment of formation of halo central part, $a_1$ is the first wave collapse time, $r_{\rm c}$ and $\rho_{\rm c}$ are the parameters of NFW fit, $L_{\rm c}$ and $\rho_{\rm m}$ are the wavelength and mean matter density at moment $a_{\rm c}$}
\label{tab1}
\begin{tabular}{|l|l|l|l|}
\hline
$A_{2,3}/A_{1}$ & $a_{\rm c}/a_1$ & $r_{\rm c}/L_{\rm c}$ & $\rho_{\rm c}/\rho_{\rm m}$ \\
\hline
0.1 & 3.4 & 0.017 & 4800  \\
0.2 & 1.9 & 0.019 & 3800  \\
0.5 & 1.0 & 0.023 & 2100  \\
1.0 & 0.64 & 0.019 & 3800  \\
\hline
\end{tabular}
\end{table}

The profile of a halo with the small scale fluctuations turned on is plotted in Figure \ref{fig_profiles} by circles. In this case it is apparent from this Figure that the density in the centre of the  halo is lower while the slope is shallower than in the case of absence of small scale fluctuations. The initial parameters are 
\be
\lambda = L/16,\quad a_{\rm s} = 0.5 a_1,
\label{small_param}
\ee
and the resulting central slope is:
\begin{equation}
\alpha = {\mathrm{d} \log \rho \over \mathrm{d} \log r} = -0.51 \pm 0.05.
\label{slope}
\end{equation}

The profile can be fitted with the following expression:
\[
 \rho = {\rho_{\rm s} \over x^{0.5}\left( 1 + x \right)^{2.5}}, \quad x = {r \over r_{\rm s}}
\]
\[
 r_{\rm s} = (0.017\pm 0.001)L_{\rm c}, \quad \rho_{\rm s} = (6.1\pm 0.2)\cdot 10^3 \rho_{\rm m}.
\]

When we take $a_{\rm s} > a_1$ the effect vanishes and the halo profile becomes almost indistinguishable from (\ref{fit}). The same happens when $\lambda \geq L/10$, $a_{\rm s} = 0.5 a_1$. For $\lambda = L/8$, $a_{\rm s} = 0.5 a_1$ the profile gets even more concentrated with $r_{\rm c}=0.007L_{\rm c}$, $\rho_{\rm c}=4.8\cdot10^4$. For  $\lambda = L/20$, $a_{\rm s} = 0.5 a_1$ the effect of flattening still exists, its amplitude, however, is smaller ($\alpha=-0.72\pm0.06$). Thus, our simulations show that the effect of flattening is maximal when $\lambda \approx L/16$.

\begin{figure}
\centering
\epsfbox{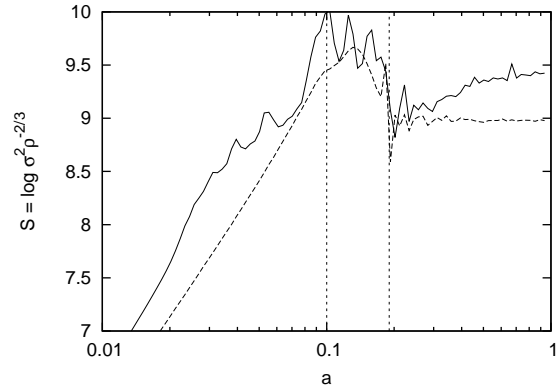} 
\caption{The evolution of the coarse grained entropy of a sphere with mass 1/20 of the mass inside $r_{\rm c}$ is shown by solid line for the simulation with small scale fluctuations turned on and by dashed line when they are turned off. Vertical lines correspond to the first wave collapse, $a_1=0.1$ and stabilization of the central part of haloes, $a_{\rm c}=0.19$.}
\label{fig_entrop}
\end{figure}

There are several possible physical effects that can take place in the system we study. One is the generation of the initial entropy. The evolution with time of the coarse grained entropy measured in spherical regions of the central parts of simulated haloes is shown in Figure \ref{fig_entrop}. The moments $a_1$ and $a_{\rm c}$ are marked by vertical lines. It is clear from this Figure that both initial (at $a<a_1$) and final ($a>a_{\rm c}$) entropy in the case of small scale fluctuations are larger by approximately the same value than in the case without small scale mode as expected from the DLM theory. The behaviour of the entropy at $a<a_1$ is well described by the initial conditions (\ref{init}), (\ref{small}). One should note that for this regime the fine-grained entropy is equal to zero while the coarse-grained one depends on the smoothing scale (but the difference between the solid and dashed lines does not depend on it when the scale is larger than $\lambda$).  The decrease of entropy in the time period corresponding to $a_1<a<a_{\rm c}$ is caused by the violent relaxation of the central regions of haloes. During this period the strong temporal and spatial variations of the gravitational potential expel a part of the material removing some energy and entropy. Finally when $a>a_{\rm c}$ some kind of thermalization occurs and the value of the coarse-grained entropy becomes independent on the way of measurement for the stable halo.

The additional small scale entropy for our toy models can be estimated following the approach of DLM. The cell size for the calculation of the coarse grained quantity $E$ should be chosen larger than the size of the small scale perturbations, $\lambda$. This means that the small scale density perturbations are averaged out (in the linear regime) and the initial entropy rises only due to the small scale velocities, so in our case of the initial conditions (\ref{init}) and (\ref{small}) we estimate:
\be
E_{\rm in} \propto \left( \Delta v_k^i\right)^2 \propto a_{\rm s}^{-2} \lambda ^2.
\label{Einit}
\ee
From equation (\ref{Einit}) it is clear that both the earlier collapse of the small scale structure and the larger $\lambda$ lead to higher entropy and therefore, may not result in a steep cusp. So the fact that the influence of the small scale perturbations decreases when $a_{\rm s} > a_1$ or $\lambda \leq L/20$ can be explained by the entropy theory.

However, $\lambda$ cannot be increased too much. One reason for this is that the simple estimate (\ref{Einit}) is valid only when $\lambda \ll L$ while for a larger $\lambda$ the effect is suppressed by the velocity correlations (see DLM). Another reason is that collapse of the small scale waves results in the formation of a large number of subhaloes, $N_{\rm sub}=(L/\lambda)^3$. For the choice of parameters (\ref{small_param}) the mass fraction in subhaloes roughly amounts to 30\% inside $r_{\rm c}$ (contrary to a few per cents in the Aquarius simulations, Springel et al. 2008). These subhaloes experience the standard two body relaxation as well as the dynamical friction (Chandrasekhar 1942; Binney \& Tremaine 1987). The duration of the relaxation process, $T_{\rm relax}$, is proportional to $N_{\rm sub}$, so when the number of subhaloes is small enough they all settle down to the centre thus increasing the central density. We observe this effect in simulations with $\lambda \geq L/10$. In these simulations subhaloes from outer parts indeed travel to the centre of the host halo and recide there. A more detailed study of the subhalo population will be given in our next paper, in preparation, as well as the impact of changing the anisotropy and small scale parameters simultaneously.

\section{A comparison with realistic simulations and observations}
In the previous Section we have shown that the simulations with two modes of perturbations demonstrate a flattening of the density cusp in the centre of a halo when the small scale mode has wavelength and collapse time in the range 
\be
1/10 < \lambda/L < 1/20,\quad a_{\rm s}<a_1.
\label{main}
\ee
One can ask whether this is enough to solve the cusp problem. Modern high resolution simulations with realistic initial conditions where one halo is simulated with $>10^9$ particles still reproduce quite well the NFW profile (Navarro et al. 2010). The more accurate fitting shows some flattening of the profile in the innermost region, $r \ll r_{\rm c}$. However the rotation curves of LSB galaxies suggest this should happen at $r\sim r_{\rm c}$, so the impact of small scales seems to be insufficient in modern simulations to solve the cusp problem.

Since the influence of small scale fluctuations depends on the ratio of amplitudes of the small and large scale modes, the effect in the real Universe should depend on the slope of the power spectrum. For the power spectrum approximated by a power law $P(k)\propto k^n$ we have checked that the conditions (\ref{main}) can be easily satisfied for $n\geq -3$, which is in agreement with the standard (not tilted) $\Lambda$CDM power spectrum at all scales. As was shown by Knollmann et al. (2008), haloes also share the NFW profile in the simulations of models having power law spectra with different slopes $n$ ranging from -0.50 to -2.75. Thus, the effect we predict has not been found in simulations with Gaussian realizations of continuous power spectrum.

In our toy models the initial conditions comprise only two modes of perturbations, while in continuous spectrum other waves may also play an important role in the formation of the density profiles of haloes. We have preliminary explored this effect with adding a third intermediate scale $\lambda<\lambda_{\rm m}<L$. We have found that the intermediate mode plays an important role in the effect of flattening of the density cusp. Considering three intermediate waves along three axes of the box, we have found that the effect of flattening vanishes when the collapse time of the intermediate mode, $a_{\rm m}$, is smaller than that of any of the large scale waves: $a_{\rm m}<{\rm min}(a_k)$. This happens because the intermediate mode produces a few subhaloes of intermediate mass. Each subhalo for its part consists only of a few smaller subsubhaloes. They have enough time to travel to the centres of subhaloes and then the main halo and steepen the cusp as has been explained in the previous Section. The number of subhaloes in the final host halo is also significantly smaller in these models. When we add the intermediate waves along only one or two axes, their impact is attenuated and cusp is suppressed.

Assuming the Gaussian initial conditions there is some probability that large scale and small scale modes satisfy the conditions (\ref{main}) while all the intermediate modes have collapse times larger than one of the large scale mode.  We estimate that this probability is about $10^{-2}$ for the slope of the power spectrum in the range $-4.0<n<-2.0$ (which corresponds to masses $M<10^{15}M_\odot$ in the standard cosmology) and it quickly decreases outside this range. 

In order to obtain this estimate we consider all waves in the range $\Delta k\sim k$ as a single mode. Therefore, for $\lambda = L/16$ we consider three large scale modes, three small scale ones and $9=3(\log_2 (L/\lambda) -1)$ intermediate modes. The estimate has the simplest form in the case of $P\propto k^{-3}$ power spectrum since all modes on average collapse simultaneously in this case. Thus, the probability for any mode to collapse earlier than some moment of time $a_1$ is $p< 1/2$ while the probability to collapse later or not collapse at all is $q=1-p$. From the results of simulations we conclude that the sufficient condition for the cusp flattening is the collapse of only 0, 1 or 2 out of 9 intermediate modes earlier than the large scale mode. The probability of such occurrence constrained by conditions (\ref{main}) is $F=p^6 q^9 + 9 p^7 q^8 + 36 p^8 q^7$. This function has its maximum $F_{\rm max}\sim 10^{-3}$ at $p\approx 1/2$. The probability of halo formation in this anzatz is determined by the probability of three modes collapsing at any time, $F_0=(1/2)^3=1/8$. The fraction of haloes without cusps is, therefore, $f=F_{\rm max}/F_0\sim 10^{-2}$. This estimate is the lower bound, its preceision, however, is about an order of magnitude. For different power spectra the calculation is somewhat more complicated since the probability $p$ for each mode depends on $k$.

The low value of the fraction of cuspless haloes in the Universe may explain why cuspless haloes have not been found in simulations with realistic initial conditions so far: they are rare. However, this estimate needs further elaborating which can be done using bulky simulations with Gaussian initial conditions.

The fraction of known galaxies without cusps is also not very high: we know less than $10^2$ such galaxies in the volume that should contain $>10^{4}$ haloes with $M>10^{11}M_\odot$ and $>10^{5}$ haloes with $M>10^{10}M_\odot$ although for the most haloes it is difficult to discriminate between cusps and cores. LSB galaxies where cores are observed are known to reside on the boundaries of voids so they may represent some special class of haloes. One way to solve the cusp problem, in our opinion, is to simulate many such low massive haloes inside a poor environment with a high resolution and to search for cores among them. Some preliminary search among the protohaloes in initial conditions can help to reduce the number of haloes to simulate.

If the discussed effect of destroying cusps due to the small scale perturbations is responsible for the cored haloes in the real Universe, it entails several predictions. First, a large number of subhaloes with masses less than $(\lambda/L)^3\sim 10^{-3}-10^{-4}$ of the main halo mass should exist in galaxies with cores. These subhaloes may change properties of a galactic disk: heat it up (Hayashi \& Chiba 2006) and produce holes in the gaseous and stellar distributions (Bekki \& Chiba 2006). Surprisingly there exist several galaxies with a large number of holes in the disk (Boomsma et al. 2008), which can form due to bombardment by such subhaloes.
These subhaloes can also be a strong source of annihilation signal if DM particles annihilate with non-zero cross-section. We estimate that the integral signal from a cored halo with subhaloes is several times higher than that of a cuspy halo with no substructure. This prediction does not make the well-known satellite over-abundance problem worse: the number of subhaloes should be higher only in some rare galaxies with cores. In cuspy haloes most of small subhaloes have been incorporated in larger subhaloes.

The effect of small scale perturbations on the halo density profile may point to another possibility to solve the cusp problem: modifications of the initial field of fluctuations such as introduction of a bump or a dip in the small scale part of the power spectrum. In such models the shape of the power spectrum guarantees the needed relation between the large, small and intermediate modes in a certain range of masses, and the probability to form a halo without cusp becomes much higher than for the standard cosmological spectrum. In particular, a spectral dip can decrease the amplitudes of the intermediate modes. Models with bump or dip were first tested by simulations in Knebe et al. (2001), however, according to our results a higher resolution is needed to see the changes of halo profiles. This models also require the exploration of the satellite over-abundance problem and the properties of the Lyman-$\alpha$ forest.

\section{Conclusions}
In this Letter we have used a set of toy models of three wave collapse to study the connection between the initial conditions and the properties of DM haloes. By changing the relative amplitudes of waves we have explored the impact of an initial anisotropy. We have found that the anisotropy affects mainly the time of halo formation and only weakly the shape of density profile of a halo.

In another set of models we have added the small scale perturbations and have found a strong flattening of the density cusp, therefore we have shown that the cusp can be flattened even for the initially cold Dark Matter. The small scale and the large scale modes must satisfy the conditions (\ref{main}) for this effect to take place. The results are in agreement with the entropy theory of DLM. The effect of flattening becomes prominent when the halo is resolved to at least $10^6$ particles. 

However, for the initial conditions with intermediate modes having $\lambda<\lambda_{\rm m}<L$ the flattening happens only when the collapse time of the intermediate mode is larger than that of any of the large scale waves. This reduces the fraction of cuspless haloes in the simulations with Gaussian initial conditions. The fraction of haloes with cores should be checked in future. If this fraction turns out to be too small to match observations, it can be increased in models with spectral bump or dip, which also need to be checked with further simulations.

The authors thank P.B. Ivanov for the discussion. The work has been supported by RFBR grants 11-02-12168, 11-02-00244 and Federal program "Scientific personnel" contract 16.740.11.0460.

\end{document}